\documentclass[preprint]{jpsj2}
\usepackage{epsf}
\usepackage{graphicx}
\catcode`\@=11
\def\eqlt{\mathrel{\mathpalette\@vereq<}}  % < over =
\def\eqgt{\mathrel{\mathpalette\@vereq>}}  % > over =
\def\@vereq#1#2{\lower2.5pt\vbox{\baselineskip0pt \lineskip-.5pt
 \ialign{$\m@th#1\hfil##\hfil$\crcr#2\crcr{=}\crcr}}}

\def\runauthor{Masatoshi {\sc Imada}
}

\title{Quantum Mott Transition and Superconductivity}  
\author{ Masatoshi  Imada$^{1,2}$ }  
\inst{$^{1}$Institute for Solid State Physics, University of Tokyo, Kashiwanoha,
Kashiwa, Chiba, 277-8581, Japan \\ 
$^2$ PRESTO, Japan Science and Technology Agency}  
\date { \today }  

\abst{
The gas-liquid transition is a first-order transition terminating at a finite-temperature critical point with diverging density fluctuations. Mott transition, a metal-insulator transition driven by Coulomb repulsion between electrons, has been identified with this textbook transition.  However, the critical temperature of the Mott transition can be suppressed, resulting in unusual quantum criticality.  It accounts for non-Fermi-liquid-like properties, and strongly momentum dependent quasiparticles as in many materials near the Mott insulator.  Among all, the mode-coupling theory of the density fluctuations supports d-wave superconductivity at the order of a hundred Kelvin for relevant parameters of the copper oxide superconductors. 
}
\kword{quantum phase transition, Mott transition, high-Tc superconductivity}

\begin{document}
\sloppy
\maketitle

\date{October 28, 2004}
%\date{October 28, 2004}
Metal-insulator transitions are characterized by a huge change of electrical conductivity.  Among all, the Mott transition is driven by the electron Coulomb repulsion~\cite{1,2}.  By using the dynamical mean-field theory~\cite{3}, Kotliar {\it et al.} suggested that this transition is equivalent to the gas-liquid transition and thus to the textbook transition of the Ising model~\cite{4,5}.  Pressure control experiment on vanadium oxides supported the equivalence~\cite{6}.  As in the gas-liquid transition, the electron density susceptibility diverges at the critical point.

Recent experimental results near the Mott insulator have clarified diversity such as renormalized Fermi liquid, charge and magnetic orders, and superconductivity with their competitions~\cite{2}.  In spectroscopic data, strongly momentum-dependent structure appears as another conspicuous feature particularly in the copper oxide superconductors~\cite{7,8}.  A well known example is the flat quasiparticle dispersions in 
the $(\pi,0)$ and $(0,\pi)$ regions of the cuprates clarified in the photoemission experiments.  Spatial inhomogeneity of electrons is also  
suggested in the cuprates~\cite{9,10} and the manganites~\cite{11,12,13}.  Severe competitions and self-organized heterogeneous structure in real and momentum spaces are a way of characterizing the region near the Mott insulator. The relation of these remarkable features to the above mentioned Mott transition but crucially modified by quantum effects was discussed recently~\cite{24}.  

In this letter we first extend the previous considerations~\cite{24} especially in two dimensions~\cite{3,6,14,15}.  The critical exponents of the Mott transition thus obtained indeed agree with the recent experimental results on a $\kappa$-(ET)$_2$-type organic compound~\cite{24-2}.
As a main result, we show that the mode coupling theory supports the d-wave superconducting state driven by this quantum Mott criticality, at the order of a hundred Kelvin for realistic parameter values of the copper oxide superconductors.

%This criticality causes the competing orders, the heterogeneous structure, 
%and non-Fermi-liquid-like properties like the linear temperature dependence of the resistivity.  
%In particular, the d-wave superconducting state is derived as inherent properties 
%near the marginal quantum critical point of the Mott transition.  

In the Ginzburg-Landau (GL) theory~\cite{16} of the gas-liquid transition, the free energy is expressed as $F=A(T)n^2/2+Bn^4/4$ with $A\propto T-T_c$, and a positive constant $B$.  The natural order parameter to be controlled is the relative density $n$ measured from the critical point. At the critical point with the temperature $T=T_c$, the density susceptibility (or compressibility) defined by $\chi_c = [d^2F/dn^2]^{-1}$ diverges. Below $T_c$, the spinodal decomposition or nucleation drive the system into a simple equilibrium with phase separation or coexistence.  

In the Mott transition, one can control either the electron filling $n$ (equivalently the hole density $X  =1- n$) or the bandwidth (or transfer) $t$ relative to the local Coulomb repulsion between electrons, $U$.  When $U/t$ increases, the conjugate variable to $U$, namely the doublon density $D$ signals the transition by discontinuous or singular reduction in $D$~\cite{3,17,18}.  Here, a doublon is defined by a site with a spin-up and a spin-down carrier electrons coexisting.  In terms of the natural order parameters, $\zeta  = X$ or $D$, the transition has a similar anomaly to that of the gas-liquid transition~\cite{4,5}. We employ the notation of the conjugate variables to $D$ and $X$ as $\mu_D\equiv U$ and the chemical potential $\mu_X\equiv\mu$, respectively. Although the full free energy may be expressed by a functional of the Green function\cite{19}, the essence is expressed by these natural order parameters after integrating out other degrees of freedom.  At $T_c$, exactly as in the gas-liquid transition, the density susceptibility (or charge compressibility) $\chi_X = [d^2F/dX^2]^{-1}$ and the doublon-density susceptibility $\chi_D = [d^2F/dD^2]^{-1}$ diverge, below which the first-order transition line with a jump of $\zeta =X$ or $D$ starts.  Since $X$ is conserved, the filling-controlled first-order transition appears as the phase separation. 

Additional factor for realistic electrons is the charge neutrality. Even when $T$ is below $T_c$ in the Hubbard-type models considering only the short-ranged force, in reality, the phase separation dynamics freezes at the stage allowed from the long-ranged repulsion, although the homogeneous phase is still unstable in the spinodal region.  This stabilizes a spatially inhomogeneous structure with the length scale determined from the balance of the spinodal instability to the electrostatic condition~\cite{24}. At $T>T_c$, if impurity potential or lattice distortions exist, it may also drive the inhomogeneity because of the enhanced density susceptibility. This reminds us of the structure observed by scanning tunnel microscope (STM) in the cuprates and manganites~\cite{9,10,11,12,13}. The present results obtained from the Mott criticality has a tight connection to the approach from dynamical stripe fluctuations~\cite{20,21}.

%
%              %%%%%%%    %%%%%        %%%%%%    %%%%%   %     %
%              %      %  %             %     %  %     %   %   %
%              %      %  %             %     %  %     %    % %
%              %%%%%%%    %%%%%        %%%%%%   %     %     %
%              %               %       %     %  %     %    % %
%              %               %       %     %  %     %   %   %
%              %         %%%%%%        %%%%%%    %%%%%   %     %
%
%              By Jean Orloff
%              Comments & suggestions by e-mail: ORLOFF@surya11.cern.ch
%              No modification of this file allowed if not e-sent to me.
%
% A simple way to measure the size of encapsulated postscript figures
%   from inside TeX, and to use it for automatically formatting texts
%   with inserted figures. Works both under Plain TeX-based macros
%   (Phyzzx, Harvmac, Psizzl, ...) and LaTeX environment.
% Provides exactly the same result on any PostScript printer provided
%   the single instruction \psfor... is changed at the end of this
%   file to fit the needs of the particular dvi->ps translator used.
%
% History:
%   1.2.4: fix error handling & add \psonlyboxes
%   1.2.3: adds \putsp@ce for OzTeX fix
%   1.2.2: makes \drawingBox \global for use in Phyzzx
%   1.2.1: accepts %%BoundingBox: (atend)
%   1.2: tries to add \psfordvitps for the TeXPS package.
%   1.1: adds \psforoztex, error handling...
%2345678 1 345678 2 345678 3 345678 4 345678 5 345678 6 345678 7 3456789
%
\catcode`\@=11
% Every macro likes a little privacy...
%
%Trying to tame the variety of \special commands for Postscript: the
%  universal internal command \PSspeci@l##1##2 takes ##1 to be the
%  filename and ##2 to be the integer scale factor*1000 (as for usual
%   TeX \scale commands)
%
\def\psfortextures{%     For TeXtures on the Macintosh
%-----------------
\def\PSspeci@l##1##2{%
\special{illustration ##1\space scaled ##2}%
}}
\def\psfordvitops{%      For the DVItoPS converter on IBM mainframes
%----------------
\def\PSspeci@l##1##2{%
\special{dvitops: import ##1\space \the\drawingwd \the\drawinght}%
}}
\def\psfordvips{%      For DVIPS converter on VAX, UNIX and PC's
%--------------
\def\PSspeci@l##1##2{%
%    \special{/@scaleunit 1000 def}% never read dox without trying!
\d@my=0.1bp \d@mx=\drawingwd \divide\d@mx by\d@my%
\includegraphics{##1\space}%
}}
\def\psforoztex{%        For the OzTeX shareware on the Macintosh
%--------------
\def\PSspeci@l##1##2{%
\special{##1 \space
      ##2 1000 div dup scale
      \putsp@ce{\number-\psllx} \putsp@ce{\number-\pslly} translate
}%
}}
\def\putsp@ce#1{#1 }
\def\psfordvitps{%       From the UNIX TeXPS package, vers.>3.12
%---------------
% Convert a dimension into the number \psn@sp (in scaled points)
\def\psdimt@n@sp##1{\d@mx=##1\relax\edef\psn@sp{\number\d@mx}}
\def\PSspeci@l##1##2{%
% psfig.psr contains the def of "startTexFig": if you can locate it
% and include the correct pathname, it should work
\special{dvitps: Include0 "psfig.psr"}% contains def of "startTexFig"
\psdimt@n@sp{\drawingwd}
\special{dvitps: Literal "\psn@sp\space"}
\psdimt@n@sp{\drawinght}
\special{dvitps: Literal "\psn@sp\space"}
\psdimt@n@sp{\psllx bp}
\special{dvitps: Literal "\psn@sp\space"}
\psdimt@n@sp{\pslly bp}
\special{dvitps: Literal "\psn@sp\space"}
\psdimt@n@sp{\psurx bp}
\special{dvitps: Literal "\psn@sp\space"}
\psdimt@n@sp{\psury bp}
\special{dvitps: Literal "\psn@sp\space startTexFig\space"}
\special{dvitps: Include1 "##1"}
\special{dvitps: Literal "endTexFig\space"}
}}
\def\psonlyboxes{%     Draft-like behaviour if none of the others works
%---------------
\def\PSspeci@l##1##2{%
\at(0cm;0cm){\boxit{\vbox to\drawinght
  {\vss
  \hbox to\drawingwd{\at(0cm;0cm){\hbox{(##1)}}\hss}
  }}}
}%
}
\def\psloc@lerr#1{%
\let\savedPSspeci@l=\PSspeci@l%
\def\PSspeci@l##1##2{%
\at(0cm;0cm){\boxit{\vbox to\drawinght
  {\vss
  \hbox to\drawingwd{\at(0cm;0cm){\hbox{(##1) #1}}\hss}
  }}}
\let\PSspeci@l=\savedPSspeci@l% restore normal output for other figs!
}%
}
%
%\def\psfor...  add your own!
%
%  \ReadPSize{PSfilename} reads the dimensions of a PostScript drawing
%	     and stores it in \drawinght(wd)
\newread\psiz@
\newdimen\drawinght\newdimen\drawingwd
\newdimen\psxoffset\newdimen\psyoffset
\newbox\drawingBox
\newif\ifNotB@undingBox
\newhelp\PShelp{Proceed: you'll have a 5cm square blank box instead of
your graphics (Jean Orloff).}
\def\@mpty{}
\def\s@tsize#1 #2 #3 #4\@ndsize{
  \def\psllx{#1}\def\pslly{#2}%
  \def\psurx{#3}\def\psury{#4}%  needed by a crazyness of dvips!
  \ifx\psurx\@mpty\NotB@undingBoxtrue% this is not a valid one!
  \else
    \drawinght=#4bp\advance\drawinght by-#2bp
    \drawingwd=#3bp\advance\drawingwd by-#1bp
%  !Units related by crazy factors as bp/pt=72.27/72 should be BANNED!
  \fi
  }
\def\sc@nline#1:#2\@ndline{\edef\p@rameter{#1}\edef\v@lue{#2}}
\def\g@bblefirstblank#1#2:{\ifx#1 \else#1\fi#2}
\def\psm@keother#1{\catcode`#112\relax}% borrowed from latex
\def\execute#1{#1}% Seems stupid, but cs are identified BEFORE execution
{\catcode`\%=12
\xdef\B@undingBox{%%BoundingBox}
}  		%% is not a true comment in PostScript, even if % is!
\def\ReadPSize#1{
 \edef\PSfilename{#1}
 \openin\psiz@=#1\relax
 \ifeof\psiz@ \errhelp=\PShelp
   \errmessage{I haven't found your postscript file (\PSfilename)}
   \psloc@lerr{was not found}
   \s@tsize 0 0 142 142\@ndsize
   \closein\psiz@
 \else
   \loop
     \execute{\begingroup
       \let\do\psm@keother
       \dospecials
       \catcode`\ =10
       \catcode`\^^M=9
       \global\read\psiz@ to\n@xtline
       \endgroup}
     \ifeof\psiz@
       \errhelp=\PShelp
       \errmessage{(\PSfilename) is not an Encapsulated PostScript File:
           I could not find any \B@undingBox: line.}
       \edef\v@lue{0 0 142 142:}
       \psloc@lerr{is not an EPSFile}
       \NotB@undingBoxfalse
     \else
       \expandafter\sc@nline\n@xtline:\@ndline
       \ifx\p@rameter\B@undingBox\NotB@undingBoxfalse
         \edef\int@rmediateresult{%
           \expandafter\g@bblefirstblank\v@lue\space\space\space}
         \expandafter\s@tsize\int@rmediateresult\@ndsize
       \else\NotB@undingBoxtrue
       \fi
     \fi
   \ifNotB@undingBox\repeat
   \closein\psiz@
 \fi
\message{#1}
}
%
% \psboxto(xdim;ydim){psfilename}: you specify the dimensions and
%    TeX uniformly scales to fit the largest one. If xdim=0pt, the
%    scale is fully determined by ydim and vice versa.
%    Notice: psboxes are a real vboxes; couldn't take hbox otherwise all
%    indentation and all cr's would be interpreted as spaces (hugh!).
%
\newcount\xscale \newcount\yscale \newdimen\pscm\pscm=1cm
\newdimen\d@mx \newdimen\d@my
\let\ps@nnotation=\relax
\def\psboxto(#1;#2)#3{\vbox{
   \ReadPSize{#3}
   \divide\drawingwd by 1000
   \divide\drawinght by 1000
   \d@mx=#1
   \ifdim\d@mx=0pt\xscale=1000
         \else \xscale=\d@mx \divide \xscale by \drawingwd\fi
   \d@my=#2
   \ifdim\d@my=0pt\yscale=1000
         \else \yscale=\d@my \divide \yscale by \drawinght\fi
   \ifnum\yscale=1000
         \else\ifnum\xscale=1000\xscale=\yscale
                    \else\ifnum\yscale<\xscale\xscale=\yscale\fi
              \fi
   \fi
   \divide \psxoffset by 1000\multiply\psxoffset by \xscale
   \divide \psyoffset by 1000\multiply\psyoffset by \xscale
   \global\divide\pscm by 1000
   \global\multiply\pscm by\xscale
   \multiply\drawingwd by\xscale \multiply\drawinght by\xscale
   \ifdim\d@mx=0pt\d@mx=\drawingwd\fi
   \ifdim\d@my=0pt\d@my=\drawinght\fi
   \message{scaled \the\xscale}
 \hbox to\d@mx{\hss\vbox to\d@my{\vss
   \global\setbox\drawingBox=\hbox to 0pt{\kern\psxoffset\vbox to 0pt{
      \kern-\psyoffset
      \PSspeci@l{\PSfilename}{\the\xscale}
      \vss}\hss\ps@nnotation}
   \global\ht\drawingBox=\the\drawinght
   \global\wd\drawingBox=\the\drawingwd
   \baselineskip=0pt
   \copy\drawingBox
 \vss}\hss}
  \global\psxoffset=0pt
  \global\psyoffset=0pt% These are local to one figure
  \global\pscm=1cm
  \global\drawingwd=\drawingwd
  \global\drawinght=\drawinght
}}
%
% \psboxscaled{scalefactor*1000}{PSfilename} allows to bypass the
%   rounding errors of TeX integer divisions for situations where the
%   TeX box should fit the original BoundingBox with a precision better
%   than 1/1000.
%
\def\psboxscaled#1#2{\vbox{
  \ReadPSize{#2}
  \xscale=#1
  \message{scaled \the\xscale}
  \divide\drawingwd by 1000\multiply\drawingwd by\xscale
  \divide\drawinght by 1000\multiply\drawinght by\xscale
  \divide \psxoffset by 1000\multiply\psxoffset by \xscale
  \divide \psyoffset by 1000\multiply\psyoffset by \xscale
  \global\divide\pscm by 1000
  \global\multiply\pscm by\xscale
  \global\setbox\drawingBox=\hbox to 0pt{\kern\psxoffset\vbox to 0pt{
     \kern-\psyoffset
     \PSspeci@l{\PSfilename}{\the\xscale}
     \vss}\hss\ps@nnotation}
  \global\ht\drawingBox=\the\drawinght
  \global\wd\drawingBox=\the\drawingwd
  \baselineskip=0pt
  \copy\drawingBox
  \global\psxoffset=0pt
  \global\psyoffset=0pt% These are local to one figure
  \global\pscm=1cm
  \global\drawingwd=\drawingwd
  \global\drawinght=\drawinght
}}
%
%  \psbox{PSfilename} makes a TeX box having the minimal size to
%      enclose the picture
\def\psbox#1{\psboxscaled{1000}{#1}}
%
% \centinsert{anybox} is just a centered \midinsert, but is included as
%    people barely use the original inserts from TeX.
%
\def\centinsert#1{\midinsert\line{\hss#1\hss}\endinsert}
\def\psannotate#1#2{\def\ps@nnotation{#2\global\let\ps@nnotation=\relax}#1}
\def\pscaption#1#2{\vbox{
   \setbox\drawingBox=#1
   \copy\drawingBox
   \vskip\baselineskip
   \vbox{\hsize=\wd\drawingBox\setbox0=\hbox{#2}
     \ifdim\wd0>\hsize
       \noindent\unhbox0\tolerance=5000
    \else\centerline{\box0}
    \fi
}}}
% for compatibility with older versions
\def\psfig#1#2#3{\pscaption{\psannotate{#1}{#2}}{#3}}
\def\psfigurebox#1#2#3{\pscaption{\psannotate{\psbox{#1}}{#2}}{#3}}
%
% \at(#1;#2)#3 puts #3 at #1-higher and #2-right of the current
%    position without moving it (to be used in annotations).
\def\at(#1;#2)#3{\setbox0=\hbox{#3}\ht0=0pt\dp0=0pt
  \rlap{\kern#1\vbox to0pt{\kern-#2\box0\vss}}}
%
% \gridfill(ht;wd) makes a 1cm*1cm grid of ht by wd whose lower-left
%   corner is the current point
\newdimen\gridht \newdimen\gridwd
\def\gridfill(#1;#2){
  \setbox0=\hbox to 1\pscm
  {\vrule height1\pscm width.4pt\leaders\hrule\hfill}
  \gridht=#1
  \divide\gridht by \ht0
  \multiply\gridht by \ht0
  \gridwd=#2
  \divide\gridwd by \wd0
  \multiply\gridwd by \wd0
  \advance \gridwd by \wd0
  \vbox to \gridht{\leaders\hbox to\gridwd{\leaders\box0\hfill}\vfill}}
%
% Useful to measure where to put annotations
\def\fillinggrid{\at(0cm;0cm){\vbox{
  \gridfill(\ht\drawingBox;\wd\drawingBox)}}}
%
% \textleftof\anybox: Sample text\endtext
%   inserts "Sample text" on the left of \anybox ie \vbox, \psbox.
%   \textrightof is the symmetric (not documented, too uggly)
% Welcome any suggestion about clean wraparound macros from
%   TeXhackers reading thisTeXhackers reading this
%
\def\textleftof#1:{
  \setbox1=#1
  \setbox0=\vbox\bgroup
    \advance\hsize by -\wd1 \advance\hsize by -2em}
\def\textrightof#1:{
  \setbox0=#1
  \setbox1=\vbox\bgroup
    \advance\hsize by -\wd0 \advance\hsize by -2em}
\def\endtext{
  \egroup
  \hbox to \hsize{\valign{\vfil##\vfil\cr%
\box0\cr%
\noalign{\hss}\box1\cr}}}
%
% \frameit{\thick}{\skip}{\anybox}
%    draws with thickness \thick a box around \anybox, leaving \skip of
%    blank around it. eg \frameit{0.5pt}{1pt}{\hbox{hello}}
% \boxit{\anybox} is a shortcut.
\def\frameit#1#2#3{\hbox{\vrule width#1\vbox{
  \hrule height#1\vskip#2\hbox{\hskip#2\vbox{#3}\hskip#2}%
        \vskip#2\hrule height#1}\vrule width#1}}
\def\boxit#1{\frameit{0.4pt}{0pt}{#1}}
\catcode`\@=12 % cs containing @ are unreachable
%
% CUSTOMIZE YOUR DEFAULT DRIVER:
%    Uncomment the line corresponding to your TeX system:
%\psfortextures%     For TeXtures on the Macintosh
%\psforoztex   %     For OzTeX shareware on the Macintosh
%\psfordvitops %     For the DVItoPS converter for TeX on IBM mainframes
 \psfordvips   %     For DVIPS converter on VAX and UNIX
%\psfordvitps  %     For dvitps from TeXPS package under UNIX
%\psonlyboxes  %     Blank Boxes (when all else fails).
%----------------------------end ofpsbox.tex & beginning of ewb.tex

\begin{figure}
$$ \psboxscaled{400}{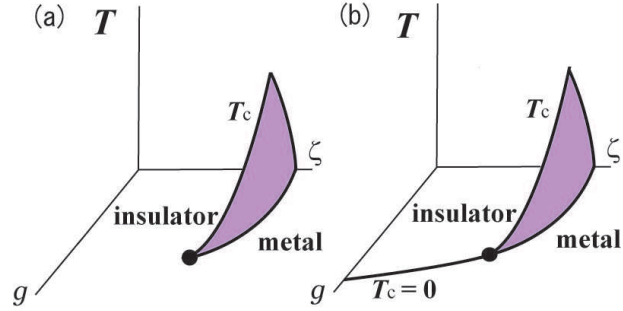} $$
\caption{Comparison of conventional quantum transition (a) with the quantum Mott transition (b) in the parameter space of $\mu_{\zeta}$, $T$ and the parameter $g$ to control quantum fluctuations. In the Mott transition (b), a zero-temperature critical line continues beyond the marginal quantum critical point (solid circle) even after the first-order transition surface (shaded sheet) shrinks.}
\label{Fig.1}
\end{figure}
How do quantum effects modify this standard classical picture? In the usual quantum critical phenomena, with increasing quantum fluctuations $g$, $T_c$ becomes lowered to zero beyond which the transition disappears as in Fig.1a.  On the contrary, the boundary between a metal and an insulator does not terminate at $T=0$. At $T=0$, metals and insulators in clean systems cannot be adiabatically connected and clearly distinguished by zero or nonzero Drude weight and density susceptibility~\cite{22,23,24}.  Therefore we have the feature in Fig.1b, in which, at $T=0$, the first-order and continuous transition lines meet at the marginal quantum critical point. 
%Furthermore, since the diverging density fluctuation coexists with the Fermi degeneracy at the marginal point, it becomes quite different from the classical transition~\cite{24}.  The diverging density susceptibility in certain cases leads to the diverging density of states $\rho(\epsilon)$. Then, instabilities towards various orders are largely enhanced and compete each other because the general form of the susceptibilities contains $\rho$. 

We now further study implications of the marginal quantum criticality.  A simple way of understanding the quantum effect is to start with the quasiparticle representation at $T=0$.  Except one-dimensional systems, the Fermi liquid theory describes metals by quasiparticles, where the interaction is renormalized to the single-particle coefficient.  In the insulating side, the single-particle Green function $G$ may also be given from a quasiparticle description with a nonzero gap $\Delta_c$ as $G(q,\omega)^{-1}=(\omega+\Delta_c+E(q,k))$ in the hole-doped region (,namely, for the pole of the lower Hubbard branch). Here, $E(q,k)$ represents the expansion around the Mott gap edge in terms of $k$ as $E(q,k)=a(q)k^2+b(q)k^4+\cdots$, where $k$ is the momentum perpendicular to the locus of Re$E(q,k)=0$ and $q$ denotes that parallel to the locus Re$E(q,k)=0$. Here Re $E(q,k)\ge 0$ should be satisfied and vanishes at the gap edge. At the transition point, $\Delta_c$ vanishes. The imaginary part of the self-energy, not considered here, does not alter the results on singularities below. 
%For continuous transitions at $T=0$, in the presence of the $a$ term, 
%the dynamical exponent z_s is given by $z_s = 2$ as in the transition to the band insulator (Refs. 2). 
If the dispersion at the Mott gap edge follows $E\propto k^{z_s}$, 
the dominant singular part of the free energy in terms of $X$ is given by $F\propto X^{(d+z_s)/d}$ and hence the susceptibility $\chi_X\propto X^{1-z_s/d}$ for the spatial dimension $d$~\cite{23,24}.  A similar scaling holds when the bandwidth is controlled by taking $D$ measured from the Mott insulator value.
%, where the overlap of upper and lower Hubbard bands 
%is caused by the control of $U/t$.  
%Since the excess doublon density is expressed by the filling 
%in the coherent band crossing. 
%the Fermi level, we have a similar form $F\propto D^{(d+z)/d}$  
%and $\chi_X\propto D^{1-z/d}$ with $D$ measured from the critical point.  
Then, for $a>0$ at $d\ge 2$, the susceptibility does not diverge, while it diverges if $a=0$ at $d\le 3$.  Along $T_c=0$ line, we expect $a>0$, if $\chi_{\zeta}$ is finite. 
In contrast, at the first-order transition as well as at the marginal critical point, the susceptibility $\chi_{\zeta}$ diverges. Therefore, when the marginal quantum critical point is approached at $T=0$ from the $T_c=0$ line, the dispersion around the Fermi level has to become flat with $a(q)=0$ at least in a part of the Brillouin zone. This alters the dynamical exponent of the single-particle excitation, $z_s$ from 2 to 4, because the $a$ term vanishes leaving quartic $b$ term as the dominant one.  The value $a$ in reality depends on momentum along the locus $E(q,k)=0$.   Near the marginal critical point, 
%is approached from the $T_c=0$ line, 
%it is unlikely that a vanishes simultaneously 
%at all the points along the $E(q,k)=0$ surface.  On the contrary, 
generically $a$ at the point with the smallest amplitude becomes zero first. Then a quartic dispersion dominates at these special points, which results in diverging $\chi_{\zeta}$.   Therefore, a singular electron differentiation generically evolves on the large Fermi surface, which consistently justifies the above scaling form of the free energy.  
This differentiation of electrons naturally explains a strong momentum dependence of quasiparticle dispersion observed in photoemission experiments with flat dispersions~\cite{7,8}.  The exponent $z_s=4$ and $\chi_X\propto X^{-1}$ were supported in several numerical calculations in two-dimensional models when filling is controlled at $T=0$, which suggests that the marginal critical point is close in these cases and a wide scaling region up to $X\sim 0.3$~\cite{2,26,27}.

Now by rescaling numerical factors, the free energy at $T=0$ is expressed as
$F= -\mu_{\zeta}\zeta + a\zeta^{(d+2)/d} +b\zeta^{(d+4)/d}+c\zeta^{(d+6)/d}\cdots$ with the constraint $\zeta\ge0$.
%Instead of the classical GL expansion;
The metal-insulator transition across the $T_c=0$ line with $a>0$ is driven by the $\mu_{\zeta}$ term.  The marginal critical point at $T=0$ may be reached at a control parameter $g=g_c$, with $a=a_0(g-g_c(T))$, $a_0>0$ and $b>0$. 
Although we consider the ``mean-field" behaviour, the $d$-dependent $F$ at $T=0$ is different from the classical GL form and gives unusually $d$-dependent exponents.
The exponent $\beta$ defined by the order parameter at $g<g_c$ and $\mu_{\zeta}=0$ as $\zeta\propto|g-g_c|^{\beta}$ is given by $\beta=d/2$.  
Near $g=g_c$, $\chi_{\zeta}$ is expressed as $\chi_{\zeta}=[d^2F/d\zeta^2]^{-1}\sim [2(d+2)a_0(g-g_c)\zeta^{2/d-1}/d^2+4(d+4)b\zeta^{4/d-1}/d^2+6(d+6)c\zeta^{6/d-1}/d^2]^{-1}$. Then $\chi_{\zeta}\propto(g-g_c)^{-\gamma}$ holds in the metallic side $g<g_c$, yielding the ``mean-field" exponent $\gamma=2-d/2$.  At $g=g_c$, $\chi_{\zeta}\sim b^{-1}\zeta^{1-\delta}$ with $\delta =4/d$ is obtained. The scaling relation $\beta\delta=\gamma+\beta$ is satisfied. 
We note that the scaling relation and the exponents are also derived from the free energy 
$F(a,\mu_{\zeta})=\xi^{-d-z_t}f(a\xi^{y_g},\mu_{\zeta}\xi^{y_{\mu}})$ with a scaling function $f$ and the correlation length $\xi\propto (g-g_c)^{-1/2}\propto \zeta^{-1/d}$, which means $y_g=2$.
Here the crossover exponent $y_{\mu}=4$ is derived from the dynamical exponent of the density (two-particle) fluctuations given by $z_t=4$.
When $T_c$ is low enough but {\it nonzero}, one may replace $g-g_c$ with $T-T_c$ in the above formalism, since $T$ may also play a role of controlling quasiparticle dispersions.
In two dimensions, it is reduced to $F=a_0(T-T_c(g))\zeta^2+b\zeta^3+c\zeta^4$ with $T_c(g_c)=0$ and 
\begin{equation}
\chi_{\zeta}=[d^2F/d\zeta^2]^{-1}\sim [2a_0(T-T_c)+6b\zeta+12c\zeta^2]^{-1}.
%\nonumber
\label{2}
\end{equation}
Then, $\gamma=1, \beta=1$ and $\delta=2$ hold~\cite{24-2}. 
Remarkably, this agrees with recent experimental result by Kagawa {\it et al.}~\cite{24-2}
%In addition to the temperature, doping and pressure dependences, 
%the exponent can also be measured by the electric-field induced transition 
%through the nonlinear conductivity.  
Very close to the critical point, the present mean-field exponent only marginally breaks down, because the Ginzburg criterion $d+z_t\ge(2\beta+\gamma)/\nu$ with $\nu=1/2$ being the correlation length exponent indicates that the system is always at the upper critical dimension.  Here we restrict ourselves to the mean field study. Although the quasi-particle picture does not hold, the one-dimensional Hubbard model with next-nearest-neighbour transfers also satisfies the above scaling at its marginal critical point, giving $\delta=4$~\cite{25}.
At $T>0$, we have the entropy term $T\zeta \ln \zeta$ in addition to the above $F$.  
This generates an essentially singular contribution with the extremum at $\zeta_0 \propto \exp[\mu_{\zeta}/T]$, which does not contribute to the present scaling behavior near $T=0$. 
Whereas at high temperatures without the Fermi degeneracy, the expansion around the extremum reproduces the regular GL form  
%\begin{equation}
$F_T=A_0(T-T_c)\zeta^2/2+B\zeta^4/4-\mu_{\zeta}\zeta$,
%\nonumber
%\end{equation}
by redefining $\zeta-\zeta_0 \rightarrow \zeta$. 

Near the critical point, the density susceptibility is enhanced.  We show consequences of the marginal quantum criticality within the mean-field scheme by assuming the dominant part of the susceptibility around a small and nonzero momentum $Q$  as
\begin{equation}
\chi_{\zeta}(q,\omega)=\frac{\Gamma^{-1}}{-i\omega+D_s(K^2+(q-Q)^2+\cdots)}.
\label{3}
\end{equation}
with a diffusion constant $D_s$, the distance from the criticality $K$ and a constant $\Gamma$. They are scaled as $\Gamma^{-1}\propto\xi^{d-4}, K\propto\xi^{-1}$ and $D_s\propto \xi^{-2}$, which reproduces the above scaling with $z_t=4$.  By using this phenomenological form, Moriya has developed the self-consistent renormalzation (SCR) theory for spin fluctuations~\cite{28}.  Now we formulate a similar scheme by considering the mode coupling for the electron-density and doublon fluctuations coming from the Mott criticality. Although we combine a perturbation treatment below, the essential ingredient of the Mott criticality may be incorporated by this scheme.
We summarize the mode coupling method for density fluctuations. The inverse susceptibility at zero frequency is written from eq.(\ref{2}) as 
\begin{equation}
\chi_{\zeta}(Q,0)^{-1} =2a_0(T-T_c)+6b\langle\zeta_Q\rangle+12c\sum_q\langle\zeta_q^2\rangle
\label{4}
\end{equation}
with the density fluctuation average
\begin{equation}
\langle\zeta_q^2\rangle=\frac{1}{2\pi}\int^{\infty}_{-\infty}d\omega \coth(\omega/2T){\rm Im}\chi_{\zeta}(q,\omega).
\label{5}
\end{equation}

\begin{figure}
$$ \psboxscaled{450}{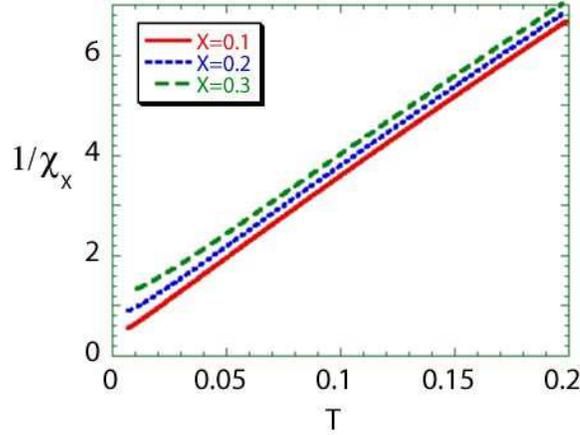} $$
\caption{The inverse of the density susceptibility peak follows a ``Curie-Weiss" temperature dependence in an extended region. For the parameter values see the text.}
\label{Fig.2}
\end{figure}
Eqs. (\ref{3})-(\ref{5}) constitute self-consistent equations to account for the Gaussian fluctuations through the mode coupling~\cite{28}. The electron self-energy is calculated perturbatively for the filling control as
\begin{equation}
\Sigma(q,\omega_n)=\frac{TU^2}{2N}\sum_{k,n}G(k,i\omega_n)\chi_{X}(q-k,i(\omega_n-\omega_m)).\label{6}
\end{equation}
In the actual calculations, though some uncertainty remains, we choose the parameter values of a two-dimensional system appropriate for the copper oxide superconductors inferred from the frequency dependence of the optical conductivity~\cite{29,30}, characteristic size of the observed inhomogeneity~\cite{9,10} and the doping dependence of the density susceptibility in numerical\cite{26} and experimental results, which give $a_0\sim0, b\sim0.7, c\sim100,\Gamma\sim3X$ and $D_s\sim30X$ by taking the energy unit $t (\sim0.4$eV) and the lattice constant as the length unit. Basically, all the results presented  here do not depend on $Q$ within the choice $0<Q<0.2\pi$.
Through the mode coupling, the Curie-Weiss type behaviour $\chi_{\zeta}\sim(T+\Theta)^{-1}$  holds in an extended temperature region with small Weiss temperature $\Theta$ near the marginal critical point (see Fig.\ref{Fig.2}) similarly to the spin fluctuation theory~\cite{28}.  Although the formalism is general, here we adopt carrier doped models with the dispersion of the square lattice $E(q)=-2t(\cos q_x+\cos q_y)$.  The imaginary part of the self-energy Im $\Sigma$ is governed by the Curie-Weiss behaviour of $\chi_X$ through eq.(\ref{6}).  Although we take $a_0=0$, the linear temperature dependence is seen in a wide temperature region.  Therefore, in the marginal-critical region with $\chi_X\propto X^{-1}$, the resistivity in two-dimensional systems becomes nearly proportional to $T$ as $\rho\propto {\rm Im} \Sigma \propto T$.  The high-energy density fluctuations typically up to around 1 eV also well explains the anomaly observed in the eV order of long-tails in the optical conductivity of many compounds near the Mott insulator~\cite{2,29,30}.

\begin{figure}
$$ \psboxscaled{450}{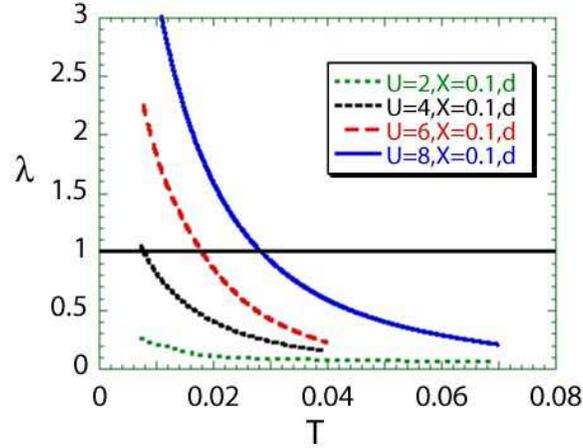} $$
\caption{The eigenvalue $\lambda$ for the linearized Eliashberg equation (7).  Among various symmetries, the eigenvalue for the $d_{x^2-y^2}$ symmetry is the largest and it is plotted as a function of temperature at $X=0.1$. The transition temperature $T_{sc}$ is given from the temperature of the eigenvalue crossing $\lambda=1$ (horizontal line).    The density susceptibility we used is the same as that in Fig.~\ref{Fig.2}.  The local interaction in the conventional Hubbard model is taken as $U$ and the nearest-neighbour transfer $t$ on the square lattice is taken as the energy unit.}
\label{Fig.3}
\end{figure}
Now we study superconductivity assuming the proximity to the Mott quantum critical point. Enhanced density fluctuations mediate the Cooper pairing.  The effective interaction between two electrons including that mediated by the density fluctuations is given by $\Lambda(q,i\omega_n)=U-U^2\chi_X(q,i\omega_n)/2$ up to the second order in $U$ with $\chi_X$ obtained from (\ref{3}).  To extract the role of density fluctuations clearly, we ignore the contribution from spin fluctuations.  The linearized Eliashberg equation for the superconducting gap $\Delta$ is given as
\begin{eqnarray}
\Delta(q,\omega_n)&=&-\frac{T}{N}\sum_{k,m}G(k,i\omega_m)G(-k,-i\omega_m) \nonumber \\
&&\times\Lambda(q-k,i(\omega_n-\omega_m))\Delta(k,i\omega_m)
\label{7}
\end{eqnarray}
for $N$-site systems.  Here we take a bare Green function ignoring the normal self-energy corrections to $G$ as the first step. In the calculation, we take the standard Hubbard model only with the nearest neighbour transfer $t$ on the square lattice and calculate the eigenvalue $\lambda$ for the right-hand side of eq.(\ref{7}).  The self-consistent solution of the linearized Eliashberg equation again for the relevant parameter values for the cuprate superconductors has the largest eigenvalue for the $d_{x^2-y^2}$  pairing symmetry and hence the highest superconducting transition temperature $T_{sc}$ ( Fig. \ref{Fig.3}) .  Even for small $Q$, the unconventional pairing comes about with the nodal structure due to the attractive part around $q=Q$ coexisting with the repulsion in the region far from $Q$ as seen in the form of $\Lambda$.  The position of the nodes with the $d_{x^2-y^2}$  symmetry is understood from the largest gap in the $(\pi,0)$ and $(0,\pi)$ regions stabilized by the flat dispersion contributing to the diverging density susceptibility.  Even for the bare Green function, the $d_{x^2-y^2}$  symmetry is stabilized by anisotropic density of states. In Fig.~\ref{Fig.3}, the transition temperature has the order of $0.01t$ to $0.05t$, which is the order of a hundred Kelvin if we take an appropriate value for the copper oxides ($t\sim 0.4$eV). 
 
In summary, we have shown that the quantum Mott criticality causes various effects including strongly momentum dependent structure of quasiparticles, non-Fermi-liquid properties and high-temperature superconductivity.
%In particular, the d-wave superconducting state is derived 
%as inherent properties near the marginal quantum critical point.  
We certainly need more elaborate treatment for better understanding on strong-coupling nature of density and doublon-density fluctuations in future steps, while all the present results indicate the significance of this proximity.  

The author is grateful to fruitful discussions with K. Kanoda, and D. N. Basov. 

\end{document}